\documentclass{emulateapj}

\shorttitle{HIP 56948: A Solar Twin}
\shortauthors{Mel\'endez \& Ram\'{\i}rez}

\begin{document}

\title {HIP 56948: A Solar Twin With a Low Lithium Abundance} 
\newcommand{\teff}{T$_{\rm eff}$ }
\newcommand{\tsin}{T$_{\rm eff}$}

% Next 5 lines define \simless and \simgreat: "less than or approximately
% equal to" and "greater than or approximately equal to".

\newbox\grsign \setbox\grsign=\hbox{$>$} \newdimen\grdimen \grdimen=\ht\grsign
\newbox\simlessbox \newbox\simgreatbox
\setbox\simgreatbox=\hbox{\raise.5ex\hbox{$>$}\llap
     {\lower.5ex\hbox{$\sim$}}}\ht1=\grdimen\dp1=0pt
\setbox\simlessbox=\hbox{\raise.5ex\hbox{$<$}\llap
     {\lower.5ex\hbox{$\sim$}}}\ht2=\grdimen\dp2=0pt\def\simgreat{\mathrel{\copy\simgreatbox}}
\def\simless{\mathrel{\copy\simlessbox}}
% Next lines define "approximately proportional to"

\newbox\simppropto
\setbox\simppropto=\hbox{\raise.5ex\hbox{$\sim$}\llap
     {\lower.5ex\hbox{$\propto$}}}\ht2=\grdimen\dp2=0pt
\def\simpropto{\mathrel{\copy\simppropto}}

\author{Jorge Mel\'endez} 
\affil{Research School of Astronomy \& Astrophysics, Mt. Stromlo Observatory, Cotter Rd., Weston Creek, ACT 2611, Australia} 
\email{jorge@mso.anu.edu.au}
\and
\author{Iv\'an Ram\'{\i}rez}
\affil{McDonald Observatory and Department of Astronomy, University of Texas at Austin, RLM 15.306 Austin TX, 78712-1083, USA}
\email{ivan@astro.as.utexas.edu} 
%\altaffiltext{1}{Also affiliated with the Seminario Permanente de Astronom\'{\i}a y Ciencias Espaciales of the Universidad Nacional Mayor de San Marcos, Peru}

%\slugcomment{Submitted to the The Astrophysical Journal Letters}
\slugcomment{Send proofs to:  J. Melendez}

\begin{abstract}
For more than a decade, 18 Sco (HD 146233) has been considered the star that most closely resembles the Sun, even though significant differences such as its Li
content, which is about three times solar, exist. Using high resolution, high S/N spectra obtained at McDonald Observatory, we show that the stars HIP~56948 and
HIP~73815 are very similar to the Sun in both stellar parameters and chemical composition, including a low Li abundance, which was previously thought to be peculiar
in the Sun. HIP~56948, in particular, has stellar parameters identical to solar within the observational uncertainties, being thus the best solar twin known to
date. HIP~56948 is also similar to the Sun in its lack of hot Jupiters. Considering the age of this star ($\sim1\pm1$\,Gyr older than the Sun) and its location and orbit around the Galaxy, if terrestrial planets exist around it, they may have had enough time to develop complex life, making it a prime target for SETI.
\end{abstract}


\section{Introduction}
Even though it is the star we know best, the Sun can not be used as the primary calibrator in stellar astrophysics because it is exceedingly bright for the instrumentation utilized for distant stars. In many cases, one would like to observe a calibrating star whose fundamental properties (temperature, luminosity, metallicity, mass and age) are identical to solar but can be analyzed in a similar way as the objects of interest.

The study of ``solar twins'' (Cayrel de Strobel 1996), stars that are spectroscopically and photometrically identical to the Sun, is necessary to set the zero point of fundamental calibrations in astrophysics, such as the color-temperature relations and the absolute flux scale (e.g., Colina et~al. 1996, Holmberg et~al. 2006). Further applications extend to many other fields. For example, in a study of the physical properties and classification of asteroids within the GAIA mission (e.g., Perryman 2005), the solar reflectance spectrum must be first removed from the asteroid observations to allow for their accurate physical modeling (A.~Korn, private communication). This can be achieved with faint solar twins, which will be found and understood more easily once the bright ones have been reliably characterized.

Solar twins can also be used to validate stellar interior and evolution models. For example, the
depletion in the observed solar Li abundance by about a factor of 200 relative to that found in
meteorites is not explained by standard stellar evolution models. Since more physically
realistic models, calculated from basic principles, cannot be computed yet, non-standard stellar
evolution models have to introduce free parameters to reproduce the observed Li
depletion (e.g., D'Antona \& Mazzitelli 1984; Ventura et~al. 1998; Charbonnel \& Talon 2005). 
Solar twins and analogs with well-determined physical and chemical properties can help validate models by providing a sample covering a wide range of Li depletion. The Sun appears to be lithium-poor by a factor of 10 compared to similar solar type disk stars (Lambert \& Reddy 2004), a fact that has led to the suggestion that the Sun is peculiar in Li and therefore of dubious value for calibrating non-standard models of Li depletion.

%Another case of apparent peculiarities in the solar abundances is that given by Ram\'{\i}rez et~al. (2007), who find that stars of solar metallicity have a super-solar oxygen abundance, a result that needs to be confirmed using stars with temperatures and metallicities very similar to solar to minimize systematic errors in the analysis. The solar oxygen abundance itself is being debated in the literature (e.g., Allende Prieto et al. 2001, Asplund et al. 2004, Mel\'endez 2004, Socas-Navarro \& Norton 2007). It is thus important to determine whether the low solar oxygen abundance is anomalous or not.

Only three solar twins are currently known: 18~Sco (Porto de Mello \& da Silva 1997), HD~98618 (Mel\'endez et~al. 2006) 
and HIP~100963 (Takeda et~al. 2007). However, these three stars have a Li abundance a factor 
of 3 to 6 higher than solar. Furthermore, according to our criteria, these stars are solar twins 
only marginally and should be called ``quasi solar twins'' instead.\footnote{Our definition of solar twin requires 
the star to have fundamental physical parameters (e.g. \tsin, log {\it g}, [Fe/H], $v_t$) identical to solar within 
a $1\sigma$ radius of observational and systematic errors. As we will see in $\S$2, HIP 56948 is the only star that can strictly
be called solar twin. Indeed, the average of six high-precision differential analyses 
(Table~3 of Mel\'endez et al. 2006, Takeda et al. 2007, our Table~1)
shows that 18 Sco, the previous best solar-twin candidate, has [Fe/H] = +0.035 ($\sigma$ = 0.010) dex,
i.e., significantly different from solar.}

Here we report the discovery of two solar twins with a low Li abundance. This study is part of a survey of solar twins and analogs in the northern hemisphere that is being completed at McDonald Observatory and will be described thoroughly in a forthcoming publication (Ram\'{\i}rez et~al. 2008).

\section{Data and Analysis}

We selected candidate solar twins from a sample of more than 100,000 stars listed in the \textit{Hipparcos} catalog by employing precise blue-optical-red-infrared color:\teff relations (Ram\'{\i}rez \& Mel\'endez 2005) with a zero-point correction based on our pilot solar-twin survey (Mel\'endez et~al. 2006), trigonometric parallaxes, and, when available, age indicators.

High resolution ($R\sim60,000$), high signal-to-noise ratio ($S/N\sim200-700$) spectra were obtained for 23 solar twin candidates and the asteroid Vesta (reflected
solar spectrum) with the 2dcoud\'e spectrograph (Tull et~al. 1995) on the 2.7-m Harlan J. Smith Telescope at McDonald Observatory in April 2007. Standard data reduction 
procedures were applied to the spectra using IRAF. Our spectra cover a wide, albeit incomplete, spectral range ($380-1000$~nm), and includes iron lines in a broad range of excitation potentials ($0-5$~eV) and line strengths. Sample spectra of the best solar twins are shown in Fig.~1.

The first analysis of the data was model-independent. Following Mel\'endez et~al. (2006), we measured the relative difference in equivalent width, $\delta W_r =
(W_\lambda^* - W_\lambda^\odot)/W_\lambda^\odot$, for a large set of \ion{Fe}{1} lines, using their median value $<\delta W_r>$ as a proxy for metallicity, while the slope in the $\delta W_r$ vs. excitation potential relation was used as a temperature indicator. Stars for which these values were zero within the observational uncertainties were considered for further detailed analysis. The results were corroborated using line-depths instead of equivalent widths. In this way we found that HIP~56948 is spectroscopically indistinguishable from the Sun and significantly a better solar twin than the other two best previously known candidates (18~Sco and HD~98618). Another close twin was identified, HIP~73815, which is, however, not as good as HIP~56948.

A detailed model-atmosphere analysis of the best candidates was performed to verify the outcome of the empirical solar-twin test and estimate precise stellar parameters, as in Mel\'endez et~al. (2006). Kurucz model atmospheres and mostly laboratory $gf$-values were employed (see Mel\'endez et al. 2006). Additionally, we measured the chromospheric emission $S$ index (Wright et al. 2004) and relative $v\sin i$ values from our spectra. The model-atmosphere analysis confirmed the empirical results, reinforcing the solar twin status of 18~Sco and HD~98618, and adding two more stars, HIP~56918 and HIP~73815, to the list of known solar twins. The parameters derived for these four solar twins are given in Table~1. HIP~56948 has solar parameters within 1-$\sigma$, while the other three candidates are solar twins within 2-$\sigma$. Relative chemical compositions are shown in Fig.~2. Except for the case of lithium (see \S3), the solar twins studied here have solar abundances within the uncertainties.

The ages of the solar twins discussed here were determined using Y$^2$ isochrones (Demarque et~al. 2004) and the chromospheric emission index $S$ (e.g., Wright et~al. 2004). Both HIP~56948 and HIP~73815 are older than the Sun by 1.2 and 1.7 Gyrs, respectively. On the other hand, 18~Sco is about 1.2~Gyrs younger than the Sun, while HD~98618 has an age similar to solar (see Table 1).

A radial velocity monitoring of what we consider to be the best solar twin ever known, HIP~56948, has been started by the McDonald Observatory planet search program (e.g., Endl et~al. 2005). Preliminary results reveal that no hot Jupiter is present around the star. Giant planets around the other two solar twins (18~Sco and HD~98618) have not been reported either by the California and Carnegie Planet Search Project (Marcy et~al. 2005). Details on the radial velocity follow up will be given in a forthcoming paper (Ram\'{\i}rez et~al. 2008).

We have also determined the Galactic space velocities and orbital parameters of the four solar twins. Like the Sun, they are in low eccentricity orbits ($e<0.1$) close to the galactic plane at a galactocentric distance of about 8\,kpc and, therefore, have an almost 100\% probability of being members of the thin disk. Furthermore, the maximum distance from the galactic plane for the orbit of HIP~56948 is only about 50\,pc.

\section{The solar twins HIP 56948 and HIP 73815}

For many years, the Sun has been thought to be peculiar in its low lithium content. This work, as well as a recent study by Takeda et al. (2007), show that this is not the case. We have found that both HIP~56948 and HIP~73815 have a low lithium abundance, similar to solar (Table 1 and Figs. 1--2), while Takeda et al. have found that some stars closely resembling the Sun also have a low Li abundance. Curiously, the other two quasi-twins (HD~98168, 18~Sco) that we have analyzed have a Li abundance about three times higher. 
HIP~100963, the other quasi-solar twin recently identified by Takeda et al. (2007), shows an even much higher Li abundance (about 6 times solar).

In order to understand the large spread in Li abundances among the five known solar twins, more high resolution, high S/N spectroscopic studies are needed to improve the precision of the
Li abundances of the Li-poor stars and explore other fragile elements such as beryllium. 
As noted by several authors (e.g., D'Antona \& Mazzitelli 1984; 
Deliyannis \& Pinsonneault 1997; Ventura et~al. 1998; Charbonnel \& Talon 2005), the amount of Li depletion 
depends on several ingredients such as mass, age, convection treatment, 
extra-mixing mechanisms (e.g., rotation), mass loss, metallicity, and magnetic fields.  
Since Li and Be are destroyed at different temperatures
($\approx$ 2.5 and 3.5 $\times 10^6$ K; Deliyannis \& Pinsonneault 1997), their observed abundances can be used to reveal the dominant mechanism(s) causing only some stars to have such low Li abundances. 

Two apparent trends in our small solar twin sample may help understand the large Li spread. 
One is that the twins with the highest masses 
(18 Sco and HD 98618, $\approx 3-4$\,\% more massive than solar) are less depleted in Li, 
in qualitative, albeit not quantitative (apparently too large Li destruction for the small mass variation), agreement with the models of Li depletion cited above. Note, however, that the solar analogs
16 Cyg A and 16 Cyg B have masses that differ by only about 2 \% (as estimated 
by our method and the relative stellar parameters determined by Deliyannis et al. 2000), yet they show very large differences (about a factor of 5) in their Li abundances (Deliyannis et al. 2000), which suggests a depletion that depends exponentially on mass. The other apparent trend is that the Li abundances in the solar twins and the Sun seem to be correlated with age, with a decrease
of about 0.1 dex in Li abundance for an increase of 0.35 Gyrs in age, in good
agreement with the predicted depletion of about 0.1 dex in Li per 0.3 Gyrs for a solar model including internal gravity waves (Fig.~2 in Charbonnel \& Talon 2005). This age correlation seems to apply
also to the 16~Cyg system, which according to our methods should be about 2--3~Gyr older than the Sun, i.e., they should have low Li abundances, and they indeed show Li abundances similar to solar (Deliyannis et~al. 2000) although 16~Cyg~B has a Li abundance about a factor of 5 lower than 16~Cyg~A.
Perhaps both (age and mass) correlations are correct, explaining thus
why 16~Cyg~A and B have low Li abundances (age effect) and at the same
time why 16~Cyg~B has much lower Li abundance than 16~Cyg~A (mass effect).
We stress here the preliminary nature of our results. Many more solar twins must be identified to confirm or reject the apparent trends with mass and/or age and help constrain stellar models of Li depletion reliably.

\section{Conclusions}

We have found two solar-twins (HIP~56948 and HIP~73815) that have a low Li abundance, similar to solar. This is remarkable given that this is a Li abundance considerably lower (by a factor of $3-6$) than that found in three previously identified solar twins (18~Sco, HD~98618, and HIP~100963). The determination of precise abundances of other fragile elements (e.g., Be) in these solar twins may lead to an understanding of the still unknown mechanism responsible for the large Li depletion in the Sun.

Neither HIP~56948, 18~Sco, nor HD~98618 have hot Jupiters. Long-term radial velocity
monitoring of these objects is encouraged, given that it could reveal, in about a
decade, whether gaseous giant planets reside within the planetary systems that these
stars may host. 

HIP~56948 has physical properties and a chemical composition identical to that of the Sun, as well as a  similar age. Stellar ages are key in the search for complex life in the universe, especially if one considers that the formation timescale of complex life is similar to that for the Earth. SETI searches can be optimized by selecting stars with ages similar to solar. Furthermore, given its orbital properties, HIP 56948 is well within the Galactic Habitable Zone (e.g., Lineweaver et~al. 2004). Considering all these facts, HIP~56948 is an excellent candidate for life outside the solar system.

The two new solar twins presented here have not been mentioned specifically as prime candidates for habitable systems in the literature, probably because ours is the first detailed spectroscopic analysis of them. Although the twin candidates are included in a catalog of nearby habitable systems (Turnbull \& Tarter 2003) based on the \textit{Hipparcos} catalog, so are thousands of other FGKM stars with parallaxes. We suggest SETI programs to give our twins top priority in their quest for intelligent life in the universe.

Altogether, the four solar twins presented in Table~1 cover the immediate past, present and future of the Sun. Long-term studies of these stars can help us understand variations of the solar activity cycle,
which are very difficult to predict theoretically. Sunspot studies cover only a few
centuries, while indirect measurements of solar activity extend to only about ten millennia 
(e.g., Usoskin et~al. 2007). Solar twins offer the opportunity to study solar activity over much longer timescales. In fact, chromospheric emission measurements of 18 Sco reveal an activity cycle of about seven years (Hall et~al. 2007),  significantly shorter than the 11-year solar cycle. In addition, these studies may help to understand anomalous variations such as the extreme low activity of the Maunder Minimum (Maunder 1890, Eddy 1976), which can be particularly interesting if they are related to climate changes on Earth (Eddy 1976, Shindell et~al. 2001). Apparently, we know no Maunder minimum star yet (Wright 2004). %High precision monitoring of their stellar activity cycles is also needed to determine their rotation periods, which will be useful to better constrain models of Li depletion (e.g., Charbonnel \& Talon 2005) as well as determine their ``rotational'' ages.

We have started photometric measurements ($uvby$ and $UBVRI$) of our solar twin sample.
The photometry of the solar twins will be employed to derive precise solar colors (the average of the colors of solar twins) and determine accurately the zero-points of the color-\teff relations, for example those of Ram\'{\i}rez \& Mel\'endez (2005), which are reliable in a relative sense: temperature differences between stars are known with good precision but the actual \teff values may be inaccurate due to potential systematic errors in the absolute flux calibration. For accurate estimates of \teff values, a zero point correction is required. Since the temperature of the Sun is known with extreme accuracy, a very well determined solar color (e.g., four solar twins can provide a solar $B-V$ color with an uncertainty of about 0.005~mag) will result in a zero point for the \teff scale as accurate as 25~K. We encourage other groups to observe these stars and extend these fundamental calibrations to as many photometric systems as possible.

Besides the astrophysical significance of solar twins, there is another motivation to look for and study them in full detail: they help answering the question of whether the Sun is unique or not (Gustafsson 1998, Gonzalez 1999), a question that has important philosophical consequences. An anomalous Sun favors some forms of the anthropic principle, those that suggest that our location in the universe is privileged for our existence as observers, but the identification of HIP~56948 as a real solar twin, although not disproving it completely, can be used as an argument against it. Also, due to its potential for hosting habitable planets, HIP~56948 may support a Copernican view for habitability in the universe.

\acknowledgements

IR would like to thank the support of the Robert~A.~Welch Foundation of Houston, Texas to David~L.~Lambert. JM acknowledges support from the Australian Research Council to Martin Asplund. We thank W.~Cochran and M.~Endl for providing the first results on the search for planets around HIP~56948 and the anonymous referee for comments that helped improve the manuscript.

%\clearpage

\begin{deluxetable}{lccccc}
\tablewidth{0pt}
\tablecolumns{5}
\scriptsize
\tablecaption{Fundamental Parameters}
\tablehead{
\colhead{Parameter (Star - Sun)}   &
\colhead{18 Sco} &
\colhead{HD 98618} &
\colhead{HIP 56948} &
\colhead{HIP 73815} &
}
\startdata
$\Delta$ v$_t$ (km s$^{-1}$) [$\pm$ 0.06]            &  +0.14   & +0.13   &  +0.01    &  +0.07    \\
$\Delta$ \teff (K)           [$\pm$ 36]           &   +57     &  +73    &  +5       &  +30      \\
$\Delta$ log $g_{\rm spec}$ (dex) [$\pm$0.05]     &  +0.01    &  -0.06  &  -0.04    & -0.08     \\
$\Delta$ log $g_{\rm Hip}$ (dex)         &  +0.01 $\pm$ 0.02  &  +0.00 $\pm$ 0.04 &  $-$0.10 $\pm$ 0.06 & $-$0.08 $\pm$ 0.06 \\
{\bf $\Delta$ log $g_{\rm adopted}$ (dex)} & +0.01 $\pm$ 0.02 &   -0.02 $\pm$ 0.03 &  $-$0.06 $\pm$ 0.04 & $-$0.08 $\pm$ 0.04 \\
{\bf $\Delta L$}\tablenotemark{a} (L$_\odot$)  & +0.06 $\pm$ 0.09 &   0.13 $\pm$ 0.11  &  0.15 $\pm$ 0.14   & 0.21 $\pm$ 0.14 \\
{[Fe/H] (dex)}  [$\pm$ 0.024]                      &  +0.04    & +0.03    &  +0.01   & +0.02 \\
{[$\alpha$/Fe]} (dex) [$\pm$ 0.030]                &   +0.01   & $-$0.01  &  +0.03   & +0.04 \\
{[Li/H]} (dex)                                     & +0.53 $\pm$ 0.08   & +0.45 $\pm$ 0.08  &  -0.02 $\pm$ 0.13 & $\lesssim -$0.15 $\pm$ 0.2 \\
$\Delta$ Mass ($M_{\odot}$)                       &   +0.04 $\pm$ 0.03    & +0.03 $\pm$ 0.03  & +0.00 $\pm$ 0.03  & +0.00 $\pm$ 0.03\\
$\Delta$ Age$_{\rm isochro}^{\rm spec}$ (Gyr) [$\pm$ 1.0]  &  $-$2.0   &  $-$0.1  &    +0.8   &  +1.8   \\
$\Delta$ Age$_{\rm isochro}^{\rm Hip}$ (Gyr) [$\pm$ 1.5]   &  $-$0.7   &  $-$0.4  &    +2.0   &  +1.0   \\
$\Delta$ Age$_{\rm chromos}$\tablenotemark{b} (Gyr) [$\pm$ 1.9]  &   +0.2    &    +1.9  &    +1.2   &  +2.2  \\
{\bf $\Delta$ Age$_{\rm adopted}$} (Gyr) [$\pm$ 1.0] &  $-$1.2 \tablenotemark{c}  &  +0.1     &    +1.2   &  +1.7   \\
$\Delta$ $S$\tablenotemark{b} (Mt. Wilson) [$\pm$ 0.02]    &  -0.002  & $-$0.020  &  $-$0.014 & $-$0.022  \\
$\Delta$ $v$ sin {\it i} (km s$^{-1}$) [$\pm$ 0.1]    &  $-$0.11  & $-$0.05  & ~0.00 & $-$0.16  \\
{distance}{(pc)}\tablenotemark{d}  & 14.0  & 38.7  & 66.6, 62.5 &  52.3  \\
\enddata
\tablenotetext{a}{Using log $g_{\rm adopted}$.}
\tablenotetext{b}{Using our data along with Duncan et~al. (1991), Wright et~al. (2004) and Hall et al. (2007) results.} 
\tablenotetext{c}{A rotational age of $-$0.6 Gyr for 18 Sco (Mel\'endez et al. 2006) was also considered.}
\tablenotetext{d}{Hipparcos; the second value given for HIP~56948 is spectroscopic.}
\label{fundamental}
\end{deluxetable}

\clearpage

\begin{figure}
\includegraphics[scale=0.85,angle=0]{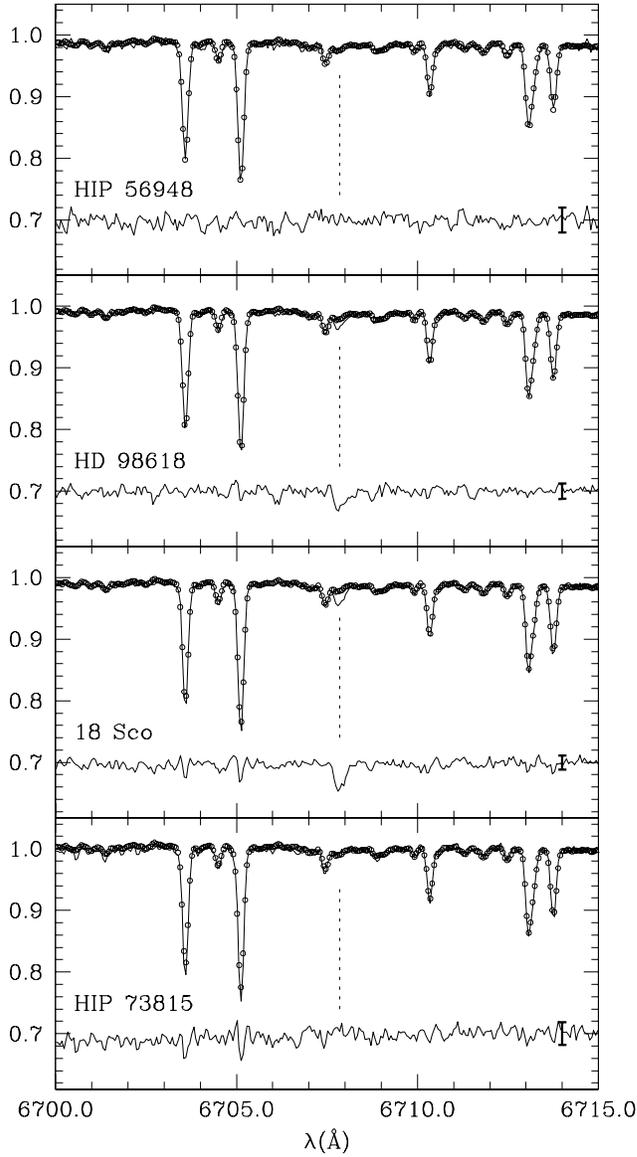}
\caption{The region around the 6708 \AA\ Li I feature (marked with vertical dotted lines) in the 
spectra of the best four solar twins (solid lines). The spectrum of Vesta (reflected solar spectrum) 
is shown with open circles (rebinned to the wavelength sampling of the solar twin spectra and slightly corrected to match their continua). Almost all the other features in this region are due to Fe I. 
Residuals (star-sun) are shown at $y=0.7$ and have been multiplied by 2 for clarity. 
The error bars at 6714 \AA\ correspond to a $2\sigma$ noise assuming that the stars are perfect 
solar twins and the noise is dominated by Poisson statistics. It is clear that the residuals for HIP~56948 are well within the noise level, including those around the Li feature. For HIP~73815 the residuals around the Li feature are consistent with the star having a solar Li abundance. The residuals for both HD~98618 and 18~Sco are not consistent with the noise level around the Li line but reasonably consistent everywhere else. The differences seen in the cores of the two strongest Fe lines for 18~Sco and HIP~73815 are due to their different $v\sin i$ values.}
\label{spectra_01} 
\end{figure}

\begin{figure}
\includegraphics[scale=0.42,angle=0]{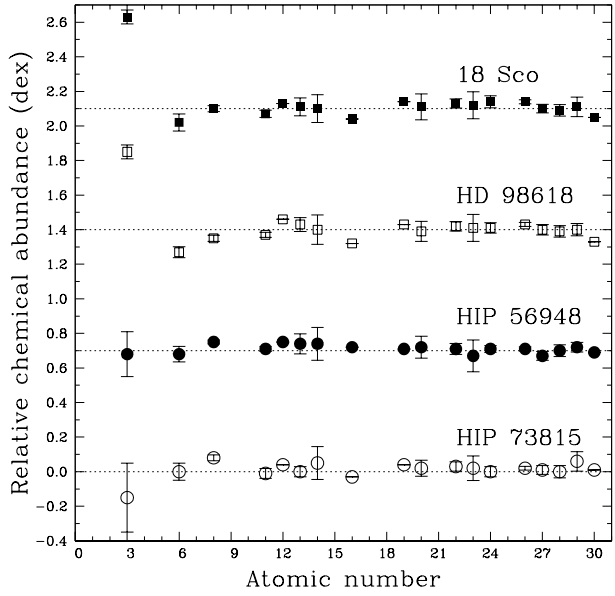} 
\caption{Chemical abundances as a function of atomic number Z. [X/Fe] is shown for all 
elements except for Li ($Z=3$) and Fe ($Z=26$), where [Li/H] and [Fe/H] are given, respectively. 
The abundances have been vertically shifted for clarity. The solar abundance pattern is represented 
by dotted lines. The error bars represent only statistical uncertainties.}
\label{spectra_02} 
\end{figure}

\end{document}